\documentclass[useAMS,usenatbib]{mn2e}

\usepackage{graphicx,amssymb,bm}
\usepackage{subfig,xfrac}
\usepackage{float, Abbrev}
\usepackage[fleqn]{amsmath}  
\usepackage{color}
\usepackage[draft]{hyperref}

\usepackage{multirow}

\newcommand{\be}{\begin{equation}}
\newcommand{\ee}{\end{equation}}
\newcommand{\ba}{\begin{eqnarray}}
\newcommand{\ea}{\end{eqnarray}}

\def\simlt{\lower.5ex\hbox{$\; \buildrel < \over \sim \;$}}

\newcommand{\fig}{\begin{figure} \begin{center}}
\newcommand{\efig}{\end{center}\end{figure} }
\newcommand{\figs}{\begin{figure*}\begin{minipage}{180mm} \begin{center}}
\newcommand{\efigs}{\end{center}\end{minipage}\end{figure*} }
\def\simgt{\lower.5ex\hbox{$\; \buildrel > \over \sim \;$}}

\usepackage{graphicx} 

\title[The shape of clusters]{The shape of relaxed galaxy clusters and the public release of a HST shape measurement code, {\sc pyRRG}}
\author[D. Harvey et al]
{David Harvey$^{1}$\thanks{e-mail: {\tt harvey@lorentz.leidenuniv.nl}}, {Sut-Ieng Tam}$^{2}$, {Mathilde Jauzac}$^{2,3,4}$, {Richard Massey}$^{2,3}$ and
\newauthor {Jason Rhodes}$^{5,6}$\\
$^{1}$Lorentz Institute, Leiden University, Niels Bohrweg 2, Leiden, NL-2333 CA, The Netherlands \\
$^{2}$Institute for Computational Cosmology, Durham University, South Road, Durham DH1 3LE, UK\\
$^{3}$Centre for Extragalactic Astronomy, Department of Physics, Durham University, Durham DH1 3LE, U.K. \\
$^{4}$Astrophysics and Cosmology Research Unit, School of Mathematical Sciences, University of KwaZulu-Natal, Durban 4041, South Africa\\
$^{5}$Jet Propulsion Laboratory, California Institute of Technology, Pasadena, CA 91109, USA\\
$^{6}$California Institute of Technology, 1201 East California Blvd., Pasadena, CA 91125, USA}
\begin{document}

\date{Accepted ---. Received ---; in original form \today.}

\pagerange{\pageref{firstpage}--\pageref{lastpage}} \pubyear{2017}

\maketitle

\label{firstpage}

\begin{abstract}
We study the shape of eight dynamically relaxed galaxy clusters observed with the {\it Hubble Space Telescope} and {\it Chandra X-Ray Observatory}. Using strong and weak gravitational lensing, the shape of the Brightest Cluster Galaxy and the X-ray isophote we study the ellipticity of the cluster halo at four different radii. We find that the proxies probing the inner regions of the cluster are strongly coupled with the BCG shape correlated with both the shape predicted by strong gravitational lensing and the X-ray isophote. Conversely we find no such correlation between the shape as predicted by the weak lensing and the other three probes suggesting any coupling between the inner and outer regions is weak. We also present in this paper the public release of the {\it HST} weak lensing shape measurement code {\sc pyRRG},  directly available from PyPi (\url{https://pypi.org/project/pyRRG/}). This {\sc python3.7} code, based on \cite{rrg} adopts an automated star-galaxy classifier based on a Random Forest and outputs scientifically useful products such as weak lensing catalogues suitable for the mass mapping algorithm {\sc Lenstool}. 
\end{abstract}


\section{Introduction}
The $\Lambda$CDM concordant model of cosmology assumes that we are living in a Universe dominated by an unknown dark energy, accelerating the expansion of space-time. This Universe is permeated by a dominant gravitating mass that we do not understand, dark matter, whilst all observable particles part of the standard model of particle physics makes up only $\sim4$\% of the energy density of the Universe \citep{planckpars,chftpars,wiggles,DEScosmology,kids450}. 

This relatively simple model of our Universe means we can describe the evolution of structure extremely well, with simulations matching observations down to $\sim1$Mpc/h \citep{Illustris,springel01,sdssboss,BAHAMAS,eagle,illustrisTNG}. The evolution of structure in this model is hierarchical, forming the smallest structures first, and mapping out a web like structure. At the nodes of this web are galaxy clusters. 

Galaxy clusters are the largest known structures in the Universe. They are dominated by dark matter, harbouring a halo of hot X-ray gas and in some cases thousands of galaxies \citep[e.g][]{Locuss_Smith,CLASH,HFF}. As extreme peaks in the density field, clusters of galaxies are ideal laboratories to study dark matter \citep[e.g.][]{Harvey_trails,Harvey_BCG,substructure_a2744A,substructure_a2744_wavelets,SIDMModel,BAHAMAS_SIDM} and constrain cosmology \citep[e.g.][]{peakCosmology,peaksnongauss,peakmodifiedGR}. 

The mass of a cluster of galaxy can exceed $10^{15}$ solar masses \citep[e.g.][]{harvey_0416,cosmicBeast}. In these environments space-time is heavily deformed, bending any geodesics that happen to pass. As such any galaxy that happens to be align itself behind the cluster with respect to our line-of-sight, will have its observed image distorted and in extreme cases stretched into arcs and split into multiple images. Modelling strong gravitational lensing has become common place when measuring the mass distribution in clusters of galaxies \citep[e.g][]{CLASH_zitrin,Merten_clash,MACSJ1149_HFF}, particularly in the core where it has a direct measure of the amount of mass. However, strong lensing has its limitations, with the constraints limited to the very core of the cluster, as such there is no information on substructures and mass in the outer regions of the cluster. Weak gravitational lensing, where the effect of the cluster must be measured statistically, grants access to this missing information \citep[e.g.][]{A1758,WtG}. It is now normal to combine both weak and strong gravitational lensing to get a full picture of the cluster \citep[e.g.][]{CLASH_zitrin,Merten_clash,strongweakunited1,strongweakunited2,MERTEN}. For a full review of mass mapping in clusters of galaxies see \cite{massModelReview}.

The exact form of galaxy clusters is still debated however it is generally accepted that they are triaxial in their shape. For a full review see \cite{clusterShapesReview}. \cite{xrayShapes} carried out a study using 25 X-ray clusters, attempting to assess the population of prolate and oblate clusters. They found that $\sim 70\%$ preferred to be prolate, however did not study the connection with other probes of the cluster environment. \cite{lensingShapes} studied the lensing properties of 25 clusters of galaxies, finding that the average ellipticity, $\langle \epsilon\rangle = 0.46\pm0.04$. They interestingly found no correlation between the lensing ellipticity and the ellipticity of the member galaxies. Moreover, they found no correlation between the position angle of the lensing signal and the cluster members, suggesting no connection.
More recently, \cite{clashMorph} carried out a study where they identified the connection between the X-ray emission, the Brightest Cluster Galaxy (BCG) and the lensing (strong and weak combined). They found that there was a strong correlation between the position angle of the cluster at large radial distances ($r\sim500$kpc), and the BCG and the inner $10$kpc, citing a coupling between the cluster and galactic star-formation properties. 
Finally a recent study of twenty relaxed and dynamically merging clusters looked at the misalignment of morphologies between the weak lensing and four probes, the Sunyaev-Zel'dovich effect, the X-ray morphology, the strong lensing morphologies and the brightest cluster galaxy \citep[BCG,][]{CLASHalignment}. 

In this paper we explicitly study the ellipticity of {\it relaxed} clusters from four separate probes; the BCG, the X-ray isophote, the strong lensing and the weak lensing. By considering each one an independent measure (including weak and strong), we can measure the connection between different regions of the cluster since each probe has a different radial dependence whilst mitigating any impact by merging structures. Moreover we present in this paper for the first time a public shape measurement code designed specifically for {\it Hubble Space Telescope (HST)}.
In section \ref{sec:lensing} we outline the basic gravitational lensing theory, in section \ref{sec:Data} we outline the data used and reduction process, in Section \ref{sec:rrg} we outline the shape measurement code {\sc pyRRG}, then in section \ref{sec:massmapp} we outline our mass mapping technique, in section \ref{sec:results} we show our results and then in section \ref{sec:conc} we conclude.

\section{Gravitational lensing}\label{sec:lensing}
Given the majority of the paper concentrates on the weak lensing of galaxy clusters, in this section we briefly outline its theoretical basics.  For a review please see \cite{BS01}, \cite{MKRev}, \cite{HoekstraRev} and \cite{RefregierRev}. For a review on strong and weak gravitational lensing please see \citep{gravitational_lensing}. 

Gravitational lensing is simply a distortion of a background source at a position, $\beta$ by foreground matter, inducing a shift of $\hat{\alpha}$ to the observed position $\theta$, i.e. 
\be
\beta = \theta - \frac{D_{\rm LS}}{D_{\rm S}}\hat{\alpha} = \theta - \alpha,
\ee
where we have introduced the reduced  deflection angle, $\alpha$. This deflection angle is related to the potential causing the deflection, which is the projected three dimensional Newtonian potential, $\Phi$,
\be
 \nabla \Psi = \alpha,
 \ee
 where 
 \be
 \Psi = \frac{D_{\rm s}}{D_{\rm L}D_{\rm S}}\frac{2}{c^2}\int\Phi(D_{\rm L},\theta,z) dz.
 \ee
 From this we can derive the distortion matrix by examining how a change in the source position effects the change in the image position, i.e. ${\partial \theta}/{\partial \beta}$, also known as the lensing Jacobian,
 \be
 A_{ij} = \frac{\partial \beta}{\partial \theta} = \delta_{ij} - \frac{\partial^2 \Psi}{\partial \theta_i\theta_j} = \delta_{ij} - \Psi_{ij},
 \ee
 where we have denoted the second derivative of $\Psi$ by the subscript, $i$ and $j$.
The second derivative of the lensing potential gives the two observables, the convergence, which is the trace of $A$,
\be
\kappa = \frac{1}{2}(\Psi_{11} + \Psi_{22}),
\ee
and corresponds to a scalar increase or decrease in the size of a distorted background source and the shear, $\gamma$ is a two component vector field given by,
\be
\gamma_1 = \frac{1}{2}(\Psi_{11} - \Psi_{22})~~~~{\rm and }~~~~ \gamma_2 = \Psi_{12} = \Psi_{21},
\ee 
corresponding to a stretch along the x-axis for $\gamma_1$ and $45^\circ$ for $\gamma_2$. We now have a relation between the observable distortion and the lensing potential. Here we limit the expansion of the Jacobian to first order, and hence assume a weak lensing limit. Indeed the shear and convergence are coupled and one cannot be observed without the other, this is known as the reduced shear,
\be
g=\gamma / (1 - \kappa).
\ee

\section{Data}\label{sec:Data}
We use {\it HST} data for both the strong and weak lensing analysis, and {\it Chandra X-Ray Observatory data} ({\it CXO}) for the X-ray analysis. We use the sample of galaxy clusters from \cite{Harvey_BCG}. This sample consists of 10 strong lensing clusters from the  Local Cluster Substructure Survey \citep[][LoCuSS]{Locuss_Richard} and the Cluster Lensing And Supernova survey with Hubble (CLASH) \citep{CLASH}. We select clusters that have at least $10$ multiple images. We place this requirement in order to ensure sufficient constraints to measure the lensing parameters that govern the inner region of the cluster. Furthermore, these ten galaxy clusters are required to be relaxed, with no signature of dynamical disturbance. We quantify this by measuring their X-ray isophotal concentration, $S=\Gamma_{\rm 100kpc} / \Gamma_{\rm 400kpc}$, where $\Gamma$ is the integrated X-ray flux within a given radius. We apply a strict criteria that the cluster must have $S>0.2$ \citep{dynamical_state_xray}. Finally of the ten clusters Abell1413 does not have sufficient optical imaging in the Advanced Camera for Surveys on {\it HST} for the weak lensing and AS1063 catalogue is currently in process from the BUFFALO survey (GO-15117) and therefore will not be ready for another year. As such we have a final sample of eight galaxy clusters.

\subsection{Strong Lensing Image Selection}
For the strong lensing measurement, we adopt the published catalogues of {\it confirmed} images, selecting those images from \cite{CLASH_zitrin}, \cite{Locuss_Richard} and \cite{a1703}.
In order to select cluster members we adopt the catalogues from  \cite{CLASH_zitrin}, \cite{Locuss_Richard} and \cite{a1703}, who had identified the red sequence in order to classify the cluster members. In order to derive the luminosities of the cluster members we match the catalogues with the photometric catalogues from \cite{CLASH_photoz}, plus any multiple image that does not have a spectroscopic redshift we use the same catalogue as a Gaussian prior with the one-sigma error as the width of the prior.
\fig
\includegraphics[width=0.5\textwidth]{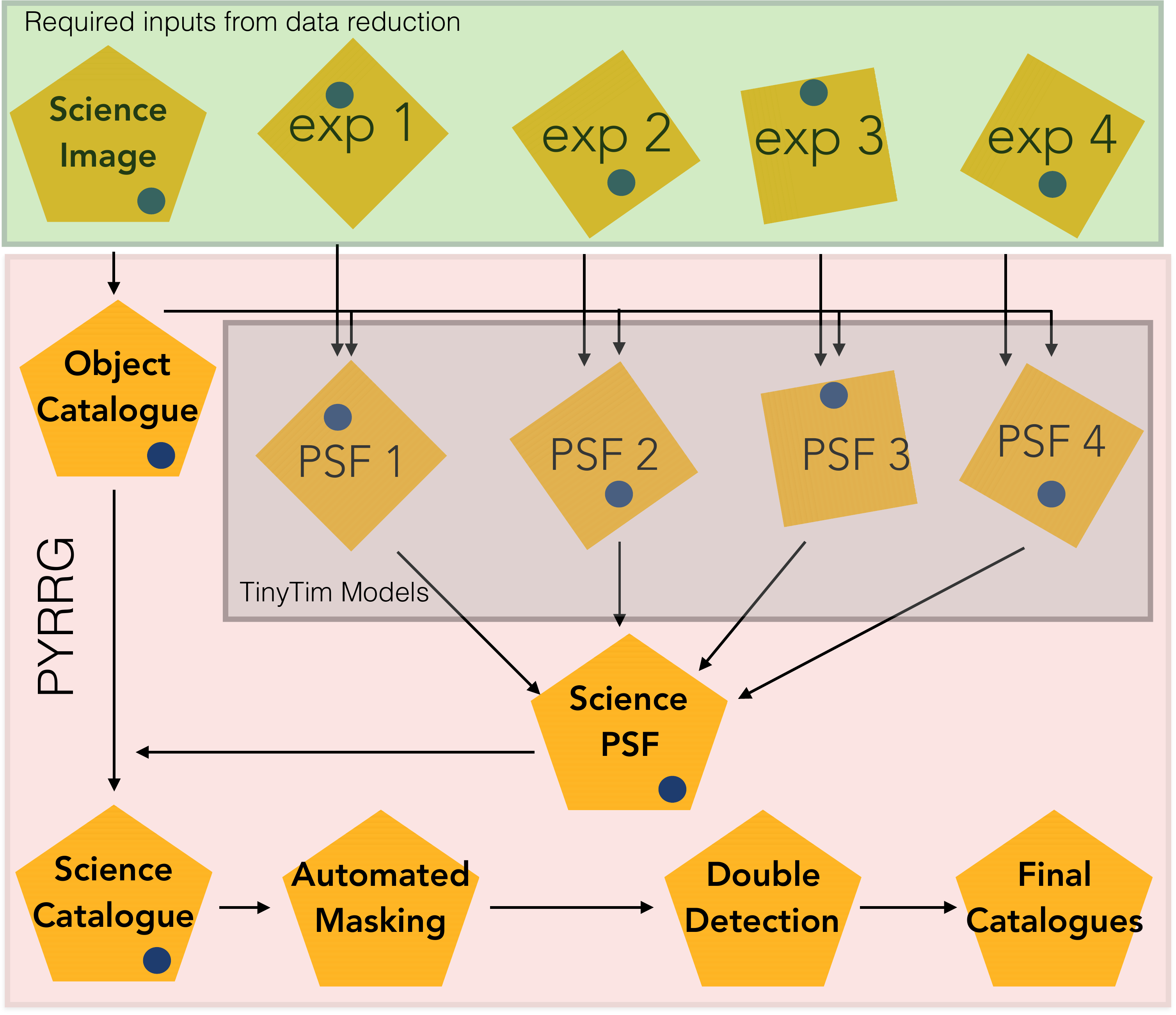}
\caption{\label{fig:PYRRG} An overview of the {\sc pyRRG} algorithm. It requires the input science image and associated weight file from the data reduction pipeline, plus all the associated exposures. From this it chooses the `best' PSF from the TinyTim models and combines them to produce a PSF at the position of each galaxy in the catalogue. It then corrects the galaxies, calculate the shears and outputs a science catalogue. It then carries out post-processing procedures to create a final `clean' catalogue.}
\efig
\subsection{Weak lensing data reduction}
We obtained raw images of each cluster with the associated reference files and re-analyse the data. We first treat each individual exposure for charge transfer inefficiency (CTI). Radiation damage from cosmic rays on the detector induces charge 'traps' in the CCD. During read-out the trapped charge is re-released intermittently causing charge to erroneously appear along the read-out axis of the CCD. As a result we model this and post-process the image in order to redistribute the charge and remove CTI `trails' \citep{CTI,CTI2,CTI3}. 

Following this we used publicly available codes to re-calibrate the raw images, {\sc CALACS}\footnote{\url{https://acstools.readthedocs.io/en/latest/calacs_hstcal.html}}, and then co-add the individual exposures, accounting for deformations induced by the telescope using the {\sc Astrodrizzle} package \citep{astrodrizzle}. During the drizzling process we use a square kernel, a pixfrac value of 0.8, and a final pixel scale of 0.03"/pixel as recommended by \cite{drizzlepars}. For those exposures that are misaligned and taken at different epochs, we use {\sc SExtractor} to extract sources from the image and then the {\sc Tweakreg} algorithm to re-align to a common reference frame. We also use {\sc Astrodrizzle} on individual exposures to produce deformation free images required by the shape measurement process, {\sc pyRRG}.

\subsection{CXO data reduction}
We re-process the raw Chandra data using the publicly available {\sc CIAO} tools\footnote{\url{http://cxc.harvard.edu/ciao/}} as in \cite{Harvey15}. We extract a region of interest and combine the exposures using the {\sc merge\_obs} script, which produces an estimate of the spatial varying PSF. We pass over a filter to remove all point sources and smooth using {\sc Wavdetect}\footnote{\url{http://cxc.harvard.edu/ciao/threads/wavdetect/}}. To extract shapes estimates, we then use {\sc SExtractor}.

\section{Shape Measurement: {\sc pyRRG}} \label{sec:rrg}
The weak lensing shape measurement consists of six key sections. An overview of the {\sc pyRRG} algorithm can be found in Figure \ref{fig:PYRRG}.
\begin{enumerate}
\item Source finding
\item Moment measuring
\item Star-Galaxy Classification
\item Point Spread Function estimation
\item Shear estimation
\item Catalogue cleaning \& masking
\end{enumerate}

\subsection{Source finding}
{\sc pyRRG} employs the `hot and cold' method that was originally developed in \cite{COSMOSintdisp} to extract sources from the image and then extended to studies used in \cite{cosmicBeast,MACSJ1149_HFF,MACSJ0717_HFF,A2744_HFF} and \cite{Harvey15_quasars}. Using the open source program {\sc SExtractor} \citep{sextractor} {\sc pyRRG} carries out two scans of the image. The first, `hot' scan, uses a smaller minimum number of pixels to count as a source, thus finding smaller objects. The second, `cold' scan, uses a larger number of pixels to classify a source. We then use the publicly available {\sc Stilts}\footnote{\url{http://www.star.bris.ac.uk/~mbt/stilts/}} software to combine the two catalogues in to one final catalogue.

\subsection{Moment measurement}
Following the source detection, {\sc pyRRG} measures the weighted multipole moments of each object in order to characterise the shape. For a full description please see \cite{rrg}, however here we outline the basics. We define the zeroth order multiple moment of a two-dimensional image in $\theta$, with an intensity distribution $i$,
\be
I=\int d^2\theta w(\theta) i(\theta), ~~~~ \label{eqn:weightedMoment}
\ee
then the quadruple {\it normalised}, weighted moment is,
\be
J_{ij}=I^{-1}\int d^2\theta~\theta_i\theta_j w(\theta) i(\theta),~~~~
\ee
followed by the second order,
\be
J_{ijk}=I^{-1}\int d^2\theta~\theta_i\theta_j\theta_k w(\theta) i(\theta),
\ee
where the weight is simply a Gaussian. From this we can define the two components of ellipticity, $\epsilon_1$ and $\epsilon_2$, as 
\be
\epsilon_1 = \frac{J_{11} - J_{22}}{J_{11} + J_{22}},~~~~~~\epsilon_2 = \frac{2J_{21}}{J_{11} + J_{22}},
\ee
and the size of the object, $d$, is given by the combination of the quadrupole moments,
\be
d=\sqrt{\frac{1}{2}(J_{11} + J_{22})}.
\ee

\subsection{Star - Galaxy Classification: Random Forest}

Following the measurement of the normalised image moments, we classify objects in to three distinct categories, stars (both saturated and not), galaxies and noise. Given that it is a simple classifying problem, we adopt a Random Forest to automatically classify this.

A Random Forest is a supervised machine learning tool that generates an ensemble of decision trees that are then trained on known data to produce predictions for unknown data. It generates a single tree by randomly subsampling the data and carrying out simple regression to create an estimator for the subsample of data. It then re-samples randomly with replacement to generate another tree. Given that each tree is a poor unbiased estimator of the truth, the aggregated estimator should be the correct one. 
The number of trees defines how good the overall estimator is, but also how long it takes to train and how large the classifier is. 

We generate a range of data to train the Random Forest. We use data from {\it HST} including a sample of $21$ SLACS galaxies, $29$ galaxy clusters, all at a range of depths. This way we try to span the entire range of parameter space including, object magnitudes, environment, and signal to noise. We generate the ground truth by manually classifying stars and galaxies from their magnitudes, $\mu_{\rm max}$, and size for individual exposures.
We then aggregate this data and parse through the Random Forest all information regarding these objects including the object magnitude, the size of the object, the brightest pixel in the object ($\mu_{\rm max}$), the second and fourth order moments, the uncorrected ellipticities, the median sky background and the variance around this, and finally the exposure time of the image. The main panel of Figure \ref{fig:classifierImportance} shows the relative importance of each feature in the classifier. The inset shows the result on a blind test galaxy, GAL-0364-52000-084. We find that the Random Forest has a $93$\% rate of correctly classifying stars, $99$\% rate of classifying galaxies and $83$\% of characterising noise (including saturated stars)\footnote{{\sc pyRRG} allows manual selection of galaxies through an interactive region selecting scheme, however the default and what is used for this work is the Random Forest}.

\fig
\includegraphics[width=0.5\textwidth]{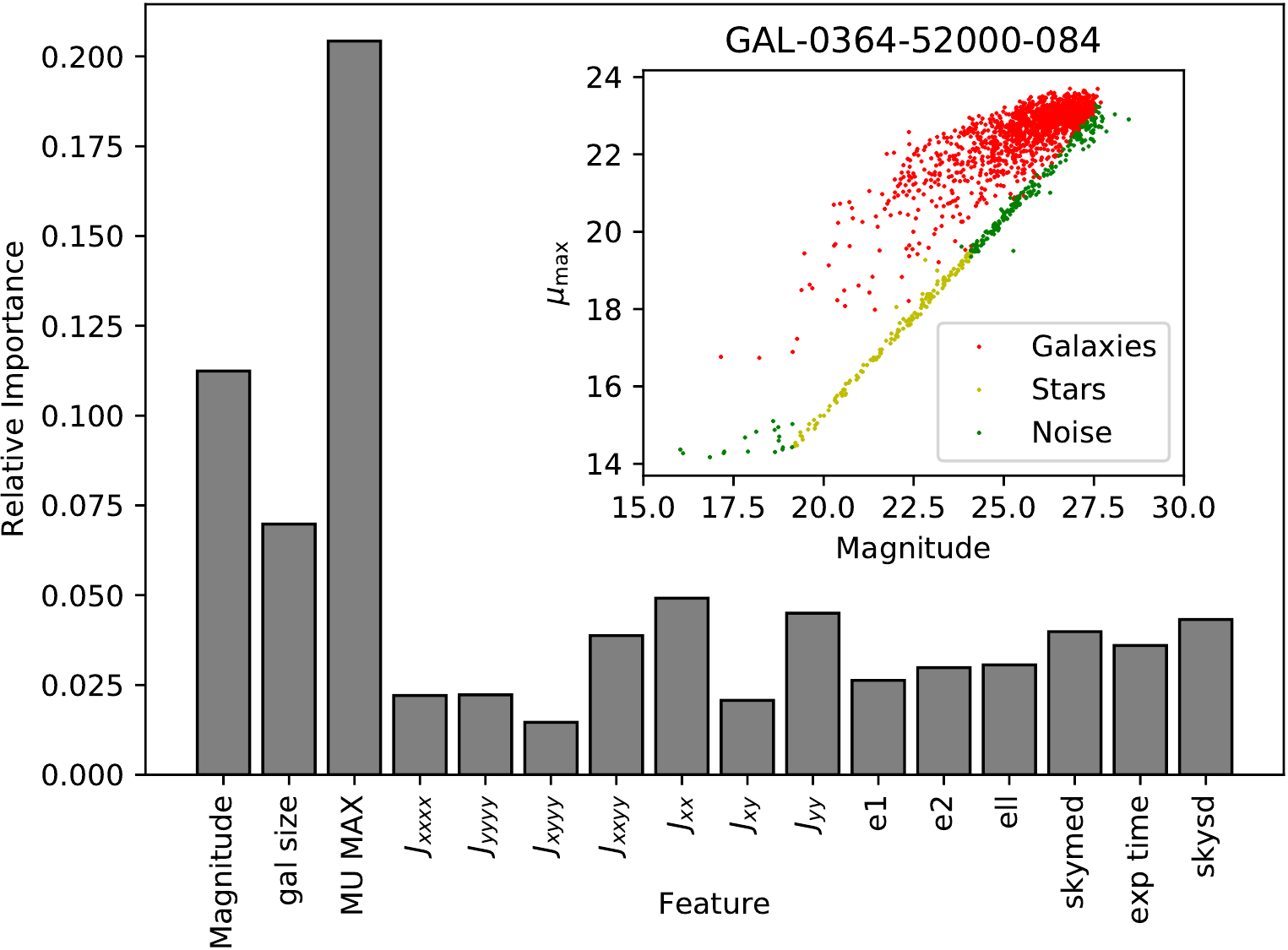}
\caption{\label{fig:classifierImportance} Results of the automated star-galaxy classifier. We use a Random Forest trained on data from the {\it HST} to predict the classification of each object. The main panel shows the relative importance of each feature in the classifier, the inset shows the results of classifying objects for the object {\it HST} field with GAL-0364-52000-084. We find that the Random Forest has a $93$\% rate of correctly classifying stars, $99$\% rate of classifying galaxies and $83$\% of characterising noise (including saturated stars).}
\efig

\subsection{Point Spread Function Measurement}

Having classified the stars, {\sc pyRRG} then estimates the impact of the telescope on the image, i.e. the Point Spread Function (PSF). HST warms up and cools down due to the heating of the Sun and therefore the focus of the telescope changes over time. We therefore estimate the impact of this by 
\begin{enumerate}
\item Taking the known positions of stars from the drizzled science image and finding the corresponding position in the individual exposures that make up that image.
\item Measuring the second and fourth order moments of the stars in each of the {\it individual exposures}
\item Having measured the moments, we compare to the various Tiny Tim models of the PSF \citep{tinytim}. We then interpolate this model to the known positions of the galaxies.
\item Combining the PSFs from each individual exposure at the position of the galaxy by rotating each PSF moment through an angle $\phi$ to the same reference frame as the drizzled science image in order to find the new rotated moment, $J'$, \citep{rotatemoments},
\begin{multline}
J'_{jk} = \sum^j_{r=0}\sum^k_{s=0}(-1)^{k-s}{j\choose k}{k\choose s}\\
\times (\cos\phi)^{j-r+s}
(\sin\phi)^{k+r-s}(J_{j+k-r-s,r+s}),
\end{multline}
and then summing the moments for a given position of the galaxy to acquire the final PSF.
\end{enumerate}
\subsection{Shear estimation}
Following the estimation of the PSF we then correct the galaxy moments and calculate the shear. 
As shown in \cite{rrg}, the final estimated shear is given by 
\be
\gamma_i = \langle\epsilon_i\rangle / G,
\ee
with
\be
G=2-\langle\epsilon^2\rangle - \frac{1}{2}\langle\lambda\rangle- \frac{1}{2}\langle\epsilon\cdot\mu\rangle,
\ee
where $\epsilon^2 = \epsilon_1^2 + \epsilon_2^2$ and  $\langle\epsilon\cdot\mu\rangle = \epsilon_1\mu_1 + \epsilon_2\mu_2$,
\be
\lambda = (J_{1111}+2J_{1122}+J_{2222})/(2d^2w^2),
\ee
and the two components of the spinor, $\mu$,
\begin{align*} 
\mu_1 &= (-J_{1111}+J_{2222})/(2d^2w^2), \\
\mu_2 &=-2(J_{1112} + J_{1222})/(2d^2w^2), \addtocounter{equation}{1}\tag{\theequation}
\end{align*}
where $w$ is the size of the weight function $w(\theta)$ in equation \eqref{eqn:weightedMoment}. From this we have a final estimator of the shear, $\gamma$.
\figs
\includegraphics[width=\textwidth]{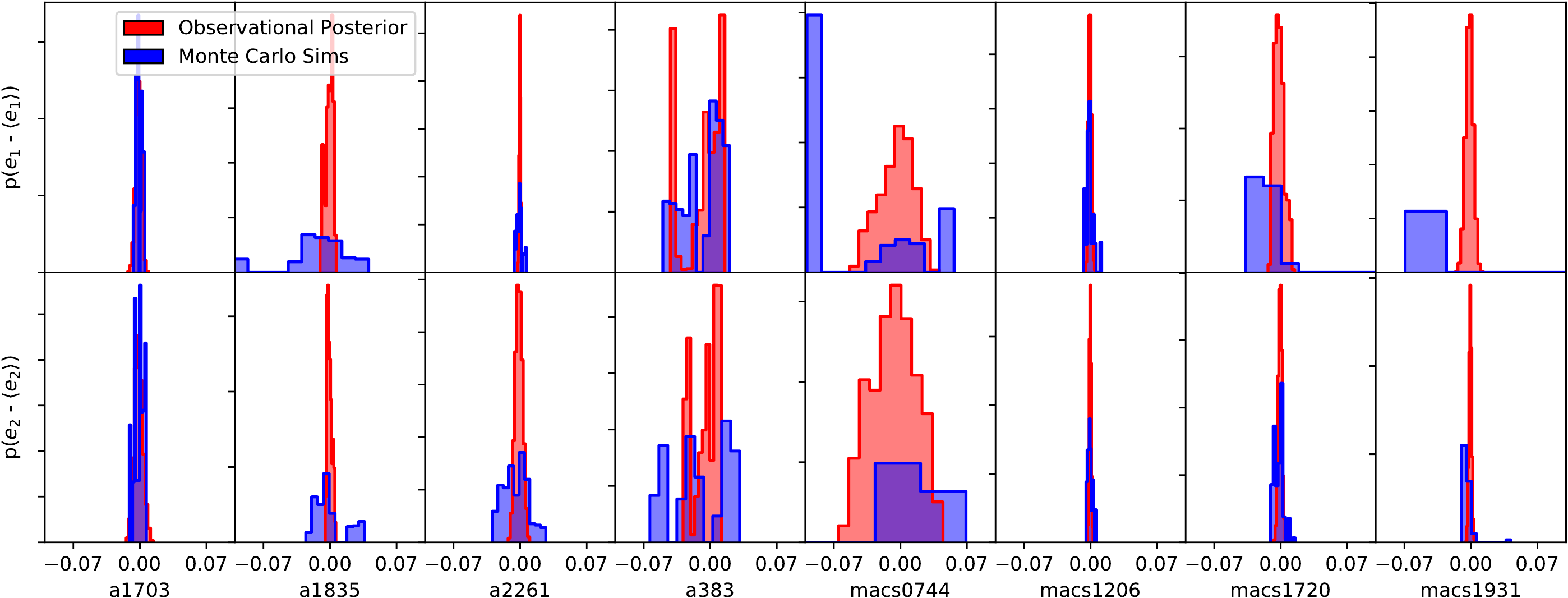}
\caption{Strong lensing error validation. We Monte Carlo the best fit lensing model 100 times. In the red is the posterior for the {\it actual} observations, and the blue histogram represents the best fit from each of the 10 Monte Carlo runs. We find that the two posteriors overlap such that the error estimates from {\sc lenstool} for strong lensing are reasonable.\label{fig:strongPosteriorWeakMonteCarloComp}}
\efigs
\figs
\includegraphics[width=\textwidth]{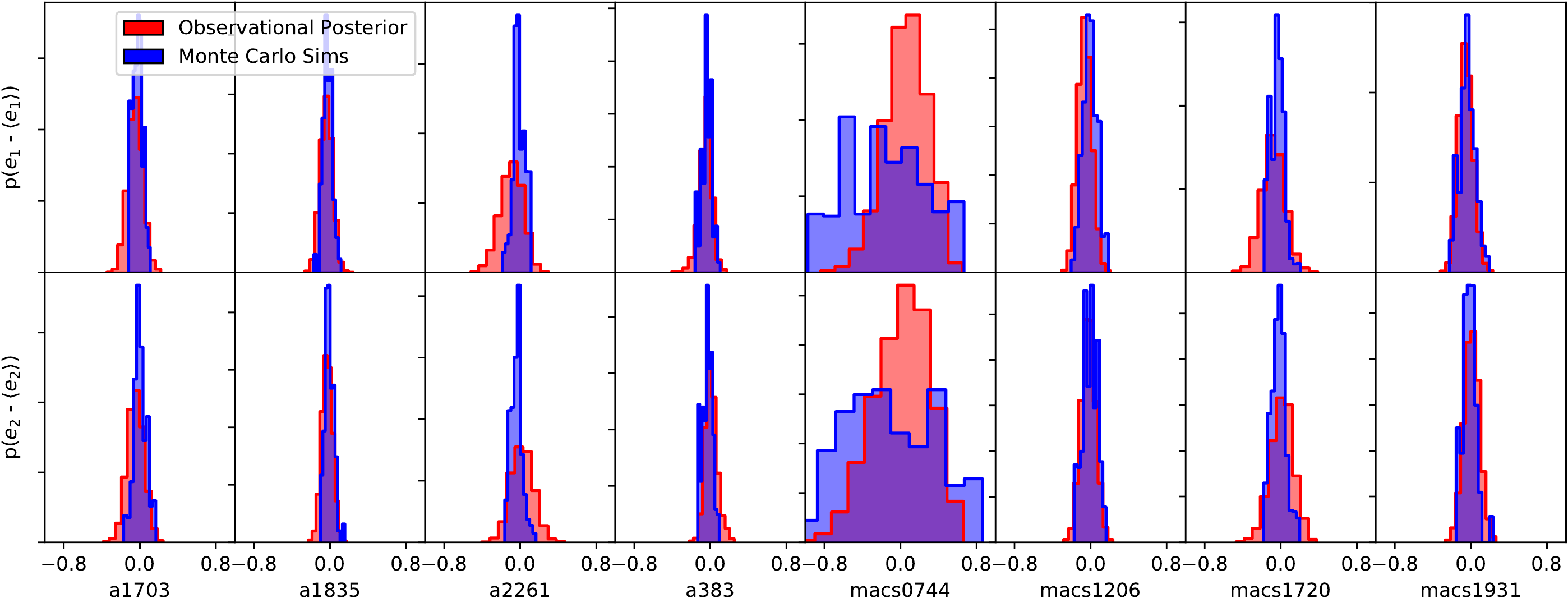}
\caption{Same as Figure \ref{fig:strongPosteriorWeakMonteCarloComp} except for weak lensing and 100 Monte Carlo runs.\label{fig:weakPosteriorWeakMonteCarloComp}}
\efigs
\subsection{ Catalogue cleaning \& masking}\label{sec:postProcess}
Having measured the shear we go through a series of cleaning operations including,
\begin{itemize}
\item{\it Automatic Masking: } Using the known position of stars and saturated stars, we generate polygons that have the same size as stars and mask any object that lies within these polygons,
\item{\it Removal of double detections: } We remove double detections whereby removing objects that lie within the isophote of a larger object.
\item{\it Creation of a {\sc Lenstool} catalogue: } Creates a catalogue to be parsed in to the mass mapping algorithm {\sc lenstool} (see Section \ref{sec:massmapp}).
\end{itemize}
Finally we match the catalogues with the CLASH catalogues, which have accurate photometric redshifts \citep{CLASH_photoz} and remove all galaxies that lie $z<z_{\rm cluster}*1.1$. This way we ensure we have no contamination, and if we remove galaxies that were very close to the cluster but still behind, their lensing signal would be very small and therefore have negligible effect on the final results. We then extract all galaxies up to 10\% beyond the strong lensing region to ensure we are not including arcs and flexed galaxies (where we define the boundary of the strong lensing region as the radius of the outermost strong lensing image).

\section{Mass mapping}\label{sec:massmapp}
We use the publicly available {\sc Lenstool} software which is a mass modelling algorithm that fits realistic parametric models to both strong and weak lensing observables to constrain the properties of the lensing potential \citep{lenstool}. It has become commonplace to use {\sc Lenstool} for combined strong and weak lensing analyses \citep[e.g.][]{MACSJ1149_HFF}, however here we want to analyse the difference in ellipticity between the core and the outer regions of the cluster. In this way we treat the strong and weak lensing reconstructions completely independently.
\figs
\includegraphics[width=\textwidth]{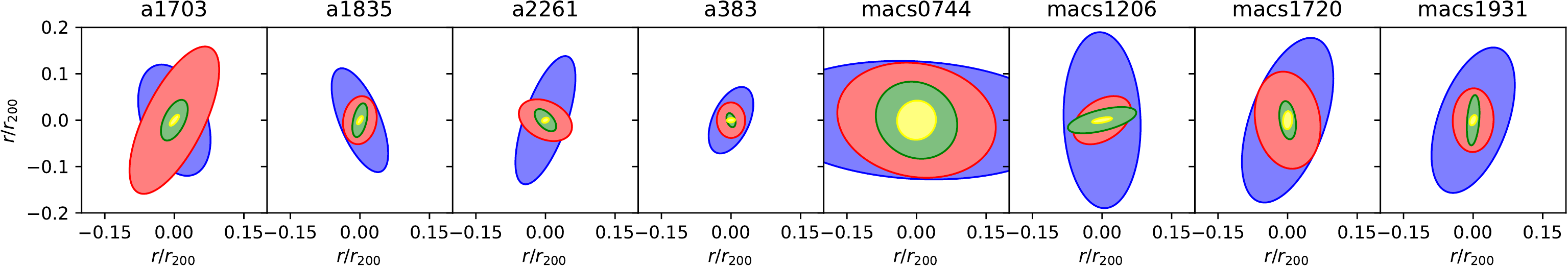}
\caption{\label{fig:strongWeakEllipses} The final results of the cluster reconstructions. We show the weak lensing, x-ray, strong lensing and BCG mass distribution in blue, red, green and yellow respectively. The mean radius of each mass component is shown as function of the virial radius. We multiple the BCG size by a factor of 10 for clarity. }
\efigs
\subsection{Strong lensing mass mapping}
We follow the same procedure as used in \cite{Harvey_BCG} whereby we choose to model the global dark matter halo with a Navarro, Frenk and White profile \citep{NFW} and model each member galaxy as a Pseudo Isothermal Elliptical Mass Distribution (PIEMD),
\be
\rho_{\rm NFW} \propto \frac{1}{x_{\rm NFW}(1.+x_{\rm NFW})^2}
\ee
and
\be
\rho_{\rm PIEMD} \propto \frac{1}{(1+x_{\rm core}^2)(1+x_{\rm cut}^2)},
\ee
where $x_{\rm NFW} = r / r_{\rm s}$, where $r_{\rm s}=r_{\rm vir}/c_{\rm vir}$,  $x_{\rm core} = r/r_{\rm core}$, $x_{\rm cut} = r/r_{\rm cut}$. We also assume that the member galaxies fall on the fundamental plane following a consistent mass to light ratio in order to reduce the parameter space such that for  the $i$th cluster member, 
\be
r_{\rm core, i} = r_{\rm core}^\star\left(\frac{L}{L^\star}\right)^{1/2},
\ee
and
\be
r_{\rm cut, i} = r_{\rm cut}^\star\left(\frac{L}{L^\star}\right)^{1/2},
\ee
and that the velocity dispersion of the galaxy is 
\be
\sigma_i = \sigma^\star\left(\frac{L}{L^\star}\right)^{1/4}.
\ee
As is common amongst strong lensing reconstructions we assume $r_{\rm core}^\star = 0.15$kpc and we have a tight Gaussian prior of $\sigma^\star=158\pm26$km/s and $r_{\rm cut}^\star=45\pm1$kpc. Following this we have $6$ free NFW parameters from the main halo, and then two free parameters for the galaxy members. In rare cases we model individual galaxies as this has shown to potentially bias mass reconstructions \citep{Harvey16}.

\subsection{Weak lensing mass mapping}
We carry out two different weak lensing reconstructions, the first is with the full catalogue and the second is with the strong lensing regime removed (see Section \ref{sec:postProcess}). We once again use {\sc Lenstool}, which estimates the weak lensing parameters by first projecting the observed ellipticities in the image plane to the source plane and then compares the subsequent source plane ellipticities with that expected from a Gaussian distribution with a width equal to the ellipticity dispersion of the sample, i.e.
\be
\chi^2 = \sum_{i=1}^2 \frac{ \epsilon_{i,s}^2}{\sigma_{\epsilon}},
\ee
where
\be
 \epsilon_{\rm s}= \frac{\epsilon - 2g + g^2\epsilon^\star}{1+|g|^2 -2\mathcal{R}(g\chi^\star)},
\ee
where we have assumed that the ellipticity can be written as a complex number in the form $\epsilon = \epsilon_1 + i\epsilon_2$, as produced from {\sc pyRRG}, and the star denotes the complex conjugate. We also adopt the mass and concentration from the strong lensing as a Gaussian prior on the weak lensing mass reconstruction.

\begin{table*}
\centering 
\begin{tabular}{|c|c|c|c|c|c|c|c|c|c|} 
 \hline 
Cluster & $M_{\rm vir}$ &  $\epsilon_{\rm 1, bcg}$ &  $\epsilon_{\rm 2, bcg}$ &  $\epsilon_{\rm 1, S}$ &    $\epsilon_{\rm 2, S}$ &    $\epsilon_{\rm 1, W}$ &    $\epsilon_{\rm 2, W}$ &    $\epsilon_{\rm 1, X}$ &    $\epsilon_{\rm 2, X}$ \\
  \hline
a1703 & 1.350 & $-0.04$ & $0.20$ & $-0.08_{0.003}^{0.003}$ & $0.10_{0.004}^{0.005}$ & $-0.06_{0.08}^{0.08}$  & $-0.05_{0.09}^{0.09}$ & $-0.13$ & $0.17$  \\
a1835 & 2.867 & $-0.07$ & $0.13$ & $-0.18_{0.005}^{0.003}$ & $0.08_{0.002}^{0.003}$ & $-0.15_{0.07}^{0.06}$  & $-0.15_{0.06}^{0.06}$ & $-0.03$ & $0.01$  \\
a2261 & 0.688 & $0.03$ & $0.03$ & $-0.01_{0.0008}^{0.0007}$ & $-0.09_{0.004}^{0.004}$ & $-0.21_{0.13}^{0.12}$  & $0.17_{0.11}^{0.12}$ & $0.02$ & $-0.03$  \\
a383 & 1.656 & $0.09$ & $0.01$ & $-0.06_{0.03}^{0.01}$ & $-0.05_{0.02}^{0.01}$ & $-0.07_{0.08}^{0.06}$  & $0.08_{0.06}^{0.07}$ & $-0.01$ & $-0.00$  \\
macs0744 & 0.995 & $-0.00$ & $0.00$ & $0.00_{0.02}^{0.02}$ & $-0.00_{0.02}^{0.02}$ & $0.19_{0.25}^{0.22}$  & $-0.02_{0.38}^{0.31}$ & $0.03$ & $-0.01$  \\
macs1206 & 1.513 & $0.36$ & $0.13$ & $0.26_{0.002}^{0.002}$ & $0.13_{0.001}^{0.001}$ & $-0.15_{0.07}^{0.07}$  & $-0.01_{0.08}^{0.08}$ & $0.02$ & $0.05$  \\
macs1720 & 0.982 & $-0.11$ & $0.02$ & $-0.16_{0.005}^{0.006}$ & $-0.05_{0.002}^{0.002}$ & $-0.12_{0.12}^{0.12}$  & $0.10_{0.13}^{0.12}$ & $-0.04$ & $-0.01$  \\
macs1931 & 0.975 & $-0.05$ & $0.05$ & $-0.37_{0.004}^{0.004}$ & $0.06_{0.001}^{0.001}$ & $-0.11_{0.07}^{0.08}$  & $0.09_{0.08}^{0.08}$ & $-0.05$ & $0.00$  \\
\hline \end{tabular} 
\caption{\label{tab:results}The best-fitting values for each of the four probes. We show in the second column the estimated virial mass of cluster from strong lensing in units of $10^{15}M_\odot$. The following eight columns show the two components of ellipticity, $\epsilon_1$ and $\epsilon_2$ for the Bright Cluster Galaxy (BCG), the strong lensing (S), the weak lensing (W) and the X-ray (X) and the given {\it statistical} error in each case.}
\end{table*}
\subsection{Ellipticity error validation}
\cite{Harvey_BCG} found that the error estimate from {\sc Lenstool} on the cartesian position of a dark matter halo using strong gravitational was underestimated by a minimum of an order of magnitude. In order to quantify whether or not the error reported using the width of the posterior distribution well reflects the true error in the ellipticity we carry out two tests; one for the strong lensing observables and one for the weak lensing. 

To do this we follow the same method as \cite{Harvey_BCG}. We mock up ten simulations based on the true data. Using the source positions from the data (for both weak and strong lensing), we use  the best fit mass model from the strong lensing reconstruction and project the sources to image positions to give a catalogue of weak and strong lensing image positions. In the case of the weak lensing we add noise through the random distribution of galaxy shapes, modelled by a Gaussian, with a mean of zero and a width equal to that of the true cluster. For the strong lensing we randomly shift the position, with the shift drawn from a Gaussian with a zero mean and a width of $0.5\arcsec$. We then reconstruct the mass distribution using {\sc Lenstool}. We Monte Carlo each cluster $100$ times for the weak lensing and $10$ for the strong, since the strong lensing reconstructions take significantly longer to converge. 
\figs
\includegraphics[width=0.49\textwidth]{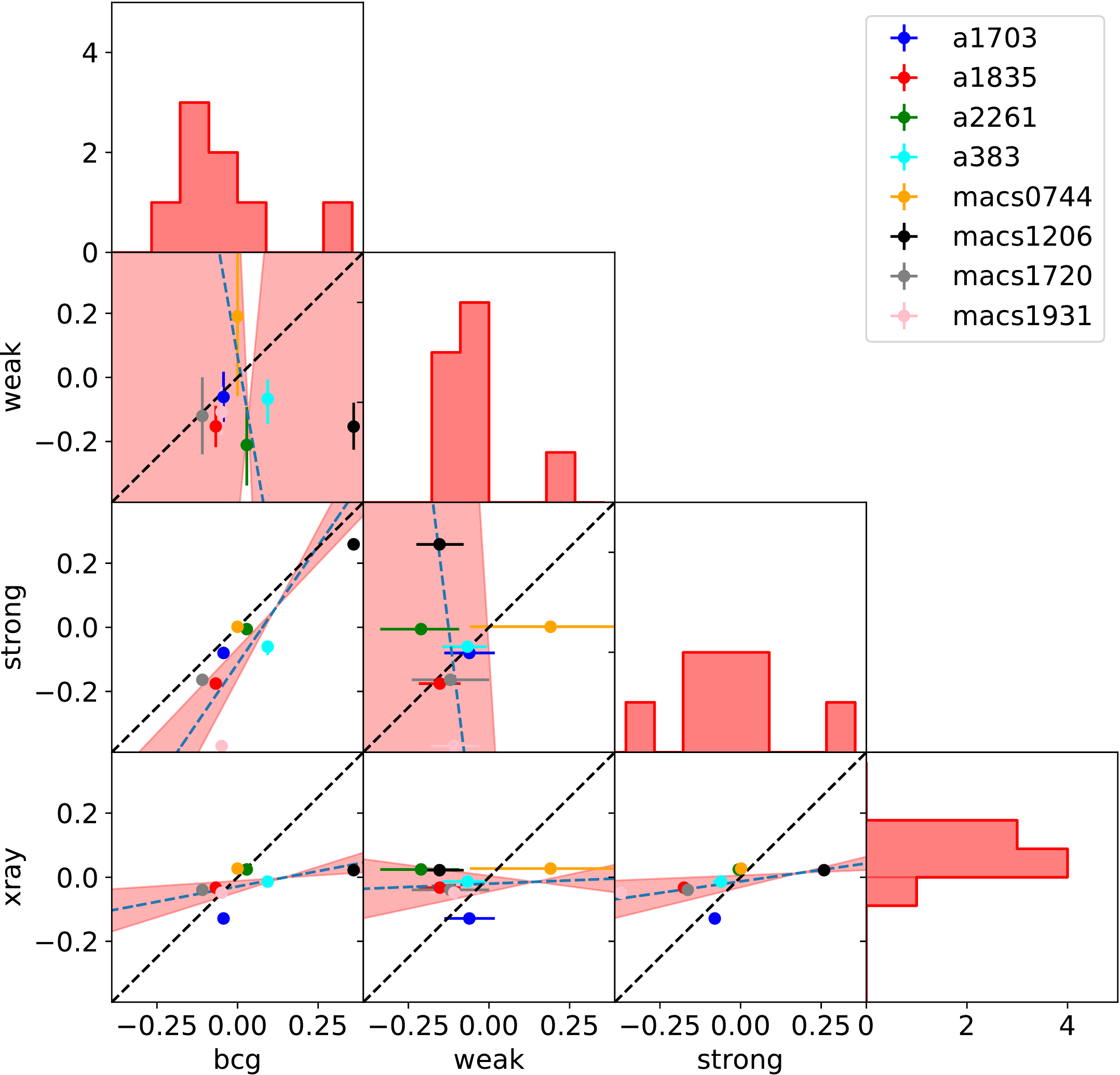}
\includegraphics[width=0.49\textwidth]{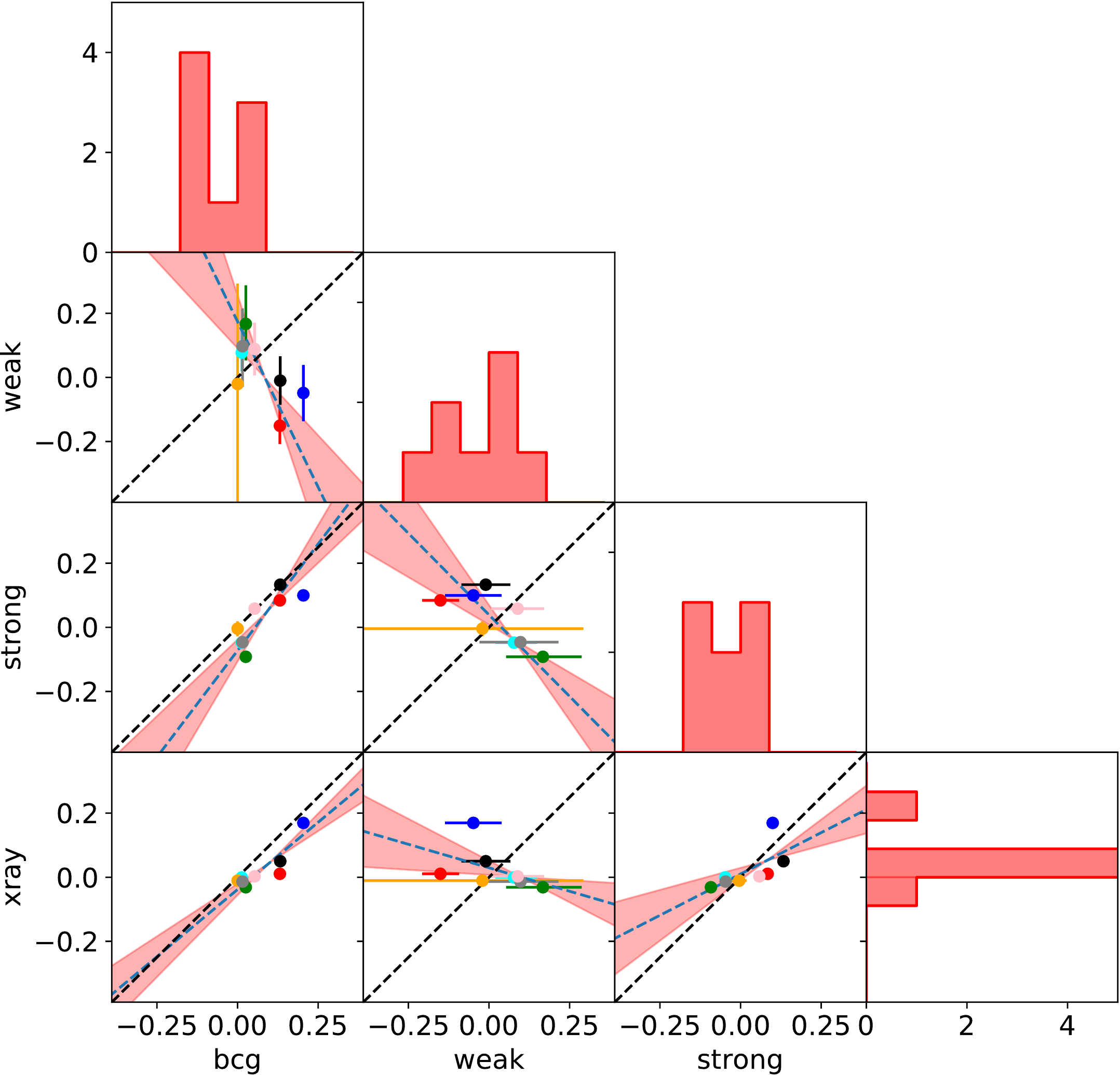}
\caption{\label{fig:BcgStrongWeak}The correlations between the different mass components. We show $e_1$ ($e_2$) on the left (right) panel. Each coloured point gives a different cluster given by the legend. We show the one to one correlation in dashed black line, and the fitted correlation in blue dashed line. We show the total histogram of each component on the diagonal of each panel.
 }
\efigs

Figure \ref{fig:strongPosteriorWeakMonteCarloComp} and Figure \ref{fig:weakPosteriorWeakMonteCarloComp} show the results of the strong and weak lensing reconstructions. Each row shows one component of ellipticity; $e_1$ and $e_2$. The red histogram is the posterior estimated by {\sc Lenstool} to the observed data, the blue histogram shows the ensemble distribution of best fitting positions from the mock simulations. Both have been shifted to zero (although in practice they have non-zero ellipticities). We find that the width of the distribution of ellipticities from the simulations closely follow the estimated posterior from the actual data. This implies that the error estimates are accurate.

\section{Results}\label{sec:results}
We carry out the strong and weak lensing mass reconstruction for the eight relaxed galaxy clusters. We show pictorially, the best fit models in Figure \ref{fig:strongWeakEllipses}. We show in blue, red, green and yellow the weak lensing, the X-ray, the strong-lensing and the BCG shape with their sizes normalised to the strong lensing estimate of the cluster virial radius. We have increased the size of the BCG by a factor of 10 for clarity. The corresponding quantitive values can be found in Table \ref{tab:results}. 

Figure \ref{fig:BcgStrongWeak} shows the correlations between each probe for each cluster, we also calculate the Pearson correlation coefficient (PCC),
\be
PCC_{1,2} = \frac{ \sum ( \epsilon_{i, 1} - \bar{\epsilon}_{i,1})(\epsilon_{i, 2} - \bar{\epsilon}_{i,2})}
{\sqrt{\sum ( \epsilon_{i, 1} - \bar{\epsilon}_{i,1})^2(\epsilon_{i, 2} - \bar{\epsilon}_{i,2})^2}},
\ee
where, $i$ corresponds to either $\epsilon_1$ or $\epsilon_2$. We show the coefficients in Table \ref{tab:pearson}. We find that unlike \cite{CLASHalignment}, there is no correlation between the weak lensing and the X-ray, and in-fact, there is some evidence for an anti-correlation, with $\epsilon_2$ showing a $PCC=-0.46$. 

We find the strongest correlation between the strong lensing and the BCG shape with a $PCC=0.83$ and $PCC=0.82$ for the two components of ellipticity. Moreover, the BCG and X-ray have an apparent correlation, with $PCC=0.50$ and $PCC=0.87$ for $\epsilon_1$ and $\epsilon_2$. We also find that the strong and weak lensing have a no $\epsilon_1$ correlation ($PCC=0.04$), however appear to have an $\epsilon_2$ anti-correlation ($PCC=-0.72$) however with such a small sample this is not significant. Similarly the weak lensing has a  $\epsilon_2$ anti-correlation ($PCC=-0.64$). Finally we find a strong correlation between the strong lensing and the X-ray emission ($PCC=0.50$ and $PCC=0.66$).

\begin{table}
\centering 
\begin{tabular}{|c|c|c|c|c|}
$\epsilon_1$ & BCG & Xray & Weak & Strong \\
 \hline 
 BCG & 1 & 0.50 & -0.14 & 0.83 \\
 Xray & 0.50 &1 & 0.10 & 0.50 \\
 Weak & -0.14 & 0.10&1 & 0.04 \\
 Strong & 0.83& 0.50 & 0.04 &1.\\
 \hline 
$\epsilon_2$  &  &  &  &  \\
 \hline 
 BCG & 1 & 0.87 & -0.64 & 0.82 \\
 Xray & 0.87 & 1 & -0.46 & 0.66 \\
 Weak & -0.64 & -0.46& 1 & -0.72 \\
 Strong & 0.82& 0.66 & -0.72 & 1 \\
 \hline 
\end{tabular} 
\caption{The Pearson correlation coefficient for the four probes for $\epsilon_1$ ({\it top}) and $\epsilon_2$ ({\it bottom}). \label{tab:pearson}}
\end{table}


\section{Discussion and conclusions}\label{sec:conc}
We have carried out an investigation in to the shape of eight dynamically {\it relaxed} galaxy clusters using a combination the {\it Hubble Space Telescope (HST)} and the {\it Chandra X-ray Observatory (CXO)}. Using a combination of Brightest Cluster Galaxy, strong gravitational lensing, X-ray emission and Weak gravitational lensing, we have measured the two components of ellipticity, $\epsilon_1$ and $\epsilon_2$ at varying radii probing different physical mechanism of a galaxy cluster. 

We found the shape of the inner regions of the cluster are strongly coupled with the BCG shape correlated with both the X-ray isophote and the strong lensing region. This implies some fundamental physics that is connecting these probes. However, we do not find any evidence for a correlation between the shape of the weak lensing regime and the inner region probes, suggesting that any coupling between these scales is weak. We consider these findings in the context of \cite{CLASHalignment} who found a correlation between the weak lensing and the X-ray where we find no such correlation. We note here two key issues when comparing our findings with that of \cite{CLASHalignment}. The first, and foremost, we remove all merging clusters or clusters that show any sign of dynamical activity in the X-ray. We therefore have a very different sample removing sub-structures that will affect the weak lensing and X-ray shape (which would have an obvious physical correlation), and secondly due to this cut we also have significantly less clusters and hence our findings are statistically weaker.

Finally, we publicly release our shape measurement code {\sc pyRRG}. Available to directly install from PyPi via \url{https://pypi.org/project/pyRRG/}, this Python3.7 code based of \cite{rrg} is specifically designed for HST shape measurement. It is fitted with an automated star-galaxy classifier and outputs scientifically useful products such as catalogues for the mass reconstruction code {\sc Lenstool}. For more see \url{https://github.com/davidharvey1986/pyRRG}.

\bsp

\section*{Acknowledgements}
DH acknowledges support by the ITP Delta foundation.
J.R. was supported by JPL, which is run under a contract for NASA by Caltech.
MJ is supported by the United Kingdom Research and Innovation (UKRI) Future Leaders Fellowship 'Using Cosmic Beasts to uncover the Nature of Dark Matter' [grant number MR/S017216/1]. This project was also supported by the Science and Technology Facilities Council [grant number ST/L00075X/1]. SIT is supported by Van Mildert College Trust PhD Scholarships. RM is supported by the Royal Society.

\label{lastpage}
\bibliographystyle{mn2e}
\bibliography{bibliography}

\begin{thebibliography}{67}
\expandafter\ifx\csname natexlab\endcsname\relax\def\natexlab#1{#1}\fi

\bibitem[{{Abbott} {et~al}\mbox{.}(2018){Abbott}, {Abdalla}, {Alarcon},
  {Aleksi{\'c}}, {Allam}, {Allen}, {Amara}, {Annis}, {Asorey}, {Avila},
  {Bacon}, {Balbinot}, {Banerji}, {Banik}, {Barkhouse}, {Baumer}, {Baxter},
  {Bechtol}, {Becker}, {Benoit-L{\'e}vy}, {Benson}, {Bernstein}, {Bertin},
  {Blazek}, {Bridle}, {Brooks}, {Brout}, {Buckley-Geer}, {Burke}, {Busha},
  {Campos}, {Capozzi}, {Carnero Rosell}, {Carrasco Kind}, {Carretero},
  {Castander}, {Cawthon}, {Chang}, {Chen}, {Childress}, {Choi}, {Conselice},
  {Crittenden}, {Crocce}, {Cunha}, {D'Andrea}, {da Costa}, {Das}, {Davis},
  {Davis}, {De Vicente}, {DePoy}, {DeRose}, {Desai}, {Diehl}, {Dietrich},
  {Dodelson}, {Doel}, {Drlica-Wagner}, {Eifler}, {Elliott}, {Elsner},
  {Elvin-Poole}, {Estrada}, {Evrard}, {Fang}, {Fernandez}, {Fert{\'e}},
  {Finley}, {Flaugher}, {Fosalba}, {Friedrich}, {Frieman},
  {Garc{\'{\i}}a-Bellido}, {Garcia-Fernandez}, {Gatti}, {Gaztanaga}, {Gerdes},
  {Giannantonio}, {Gill}, {Glazebrook}, {Goldstein}, {Gruen}, {Gruendl},
  {Gschwend}, {Gutierrez}, {Hamilton}, {Hartley}, {Hinton}, {Honscheid},
  {Hoyle}, {Huterer}, {Jain}, {James}, {Jarvis}, {Jeltema}, {Johnson},
  {Johnson}, {Kacprzak}, {Kent}, {Kim}, {King}, {Kirk}, {Kokron}, {Kovacs},
  {Krause}, {Krawiec}, {Kremin}, {Kuehn}, {Kuhlmann}, {Kuropatkin}, {Lacasa},
  {Lahav}, {Li}, {Liddle}, {Lidman}, {Lima}, {Lin}, {MacCrann}, {Maia},
  {Makler}, {Manera}, {March}, {Marshall}, {Martini}, {McMahon}, {Melchior},
  {Menanteau}, {Miquel}, {Miranda}, {Mudd}, {Muir}, {M{\"o}ller}, {Neilsen},
  {Nichol}, {Nord}, {Nugent}, {Ogando}, {Palmese}, {Peacock}, {Peiris},
  {Peoples}, {Percival}, {Petravick}, {Plazas}, {Porredon}, {Prat}, {Pujol},
  {Rau}, {Refregier}, {Ricker}, {Roe}, {Rollins}, {Romer}, {Roodman},
  {Rosenfeld}, {Ross}, {Rozo}, {Rykoff}, {Sako}, {Salvador}, {Samuroff},
  {S{\'a}nchez}, {Sanchez}, {Santiago}, {Scarpine}, {Schindler}, {Scolnic},
  {Secco}, {Serrano}, {Sevilla-Noarbe}, {Sheldon}, {Smith}, {Smith}, {Smith},
  {Soares-Santos}, {Sobreira}, {Suchyta}, {Tarle}, {Thomas}, {Troxel},
  {Tucker}, {Tucker}, {Uddin}, {Varga}, {Vielzeuf}, {Vikram}, {Vivas},
  {Walker}, {Wang}, {Wechsler}, {Weller}, {Wester}, {Wolf}, {Yanny}, {Yuan},
  {Zenteno}, {Zhang}, {Zhang}, {Zuntz}, \& {Dark Energy Survey
  Collaboration}}]{DEScosmology}
{Abbott} T.~M.~C. {et~al.}, 2018, \prd, 98, 043526

\bibitem[{{Anderson} \& {Bedin}(2010)}]{CTI3}
{Anderson} J., {Bedin} L.~R., 2010, \pasp, 122, 1035

\bibitem[{{Bartelmann}(2010)}]{gravitational_lensing}
{Bartelmann} M., 2010, Classical and Quantum Gravity, 27, 233001

\bibitem[{{Bartelmann} \& {Schneider}(2001)}]{BS01}
{Bartelmann} M., {Schneider} P., 2001, \physrep, 340, 291

\bibitem[{{Bertin} \& {Arnouts}(1996)}]{sextractor}
{Bertin} E., {Arnouts} S., 1996, \aaps, 117, 393

\bibitem[{{Cacciato} {et~al}\mbox{.}(2006){Cacciato}, {Bartelmann},
  {Meneghetti}, \& {Moscardini}}]{strongweakunited1}
{Cacciato} M., {Bartelmann} M., {Meneghetti} M., {Moscardini} L., 2006, \aap,
  458, 349

\bibitem[{{Cardone} {et~al}\mbox{.}(2013){Cardone}, {Camera}, {Mainini},
  {Romano}, {Diaferio}, {Maoli}, \& {Scaramella}}]{peakmodifiedGR}
{Cardone} V.~F., {Camera} S., {Mainini} R., {Romano} A., {Diaferio} A., {Maoli}
  R., {Scaramella} R., 2013, \mnras, 430, 2896

\bibitem[{{Diego} {et~al}\mbox{.}(2007){Diego}, {Tegmark}, {Protopapas}, \&
  {Sandvik}}]{strongweakunited2}
{Diego} J.~M., {Tegmark} M., {Protopapas} P., {Sandvik} H.~B., 2007, \mnras,
  375, 958

\bibitem[{{Donahue} {et~al}\mbox{.}(2016){Donahue}, {Ettori}, {Rasia},
  {Sayers}, {Zitrin}, {Meneghetti}, {Voit}, {Golwala}, {Czakon}, {Yepes},
  {Baldi}, {Koekemoer}, \& {Postman}}]{clashMorph}
{Donahue} M. {et~al.}, 2016, \apj, 819, 36

\bibitem[{{Harvey} \& {Courbin}(2015)}]{Harvey15_quasars}
{Harvey} D., {Courbin} F., 2015, \mnras, 451, L95

\bibitem[{{Harvey} {et~al}\mbox{.}(2017{\natexlab{a}}){Harvey}, {Courbin},
  {Kneib}, \& {McCarthy}}]{Harvey_BCG}
{Harvey} D., {Courbin} F., {Kneib} J.~P., {McCarthy} I.~G., 2017{\natexlab{a}},
  ArXiv: 1703.07365

\bibitem[{{Harvey} {et~al}\mbox{.}(2016){Harvey}, {Kneib}, \&
  {Jauzac}}]{Harvey16}
{Harvey} D., {Kneib} J.~P., {Jauzac} M., 2016, \mnras, 458, 660

\bibitem[{Harvey {et~al}\mbox{.}(2015)Harvey, Massey, Kitching, Taylor, \&
  Tittley}]{Harvey15}
Harvey D., Massey R., Kitching T., Taylor A., Tittley E., 2015, Science, 347,
  1462

\bibitem[{{Harvey} {et~al}\mbox{.}(2017{\natexlab{b}}){Harvey}, {Robertson},
  {Massey}, \& {Kneib}}]{Harvey_trails}
{Harvey} D., {Robertson} A., {Massey} R., {Kneib} J.-P., 2017{\natexlab{b}},
  \mnras, 464, 3991

\bibitem[{{Hildebrandt} {et~al}\mbox{.}(2017){Hildebrandt}, {Viola}, {Heymans},
  {Joudaki}, {Kuijken}, {Blake}, {Erben}, {Joachimi}, {Klaes}, {Miller},
  {Morrison}, {Nakajima}, {Verdoes Kleijn}, {Amon}, {Choi}, {Covone}, {de
  Jong}, {Dvornik}, {Fenech Conti}, {Grado}, {Harnois-D{\'e}raps}, {Herbonnet},
  {Hoekstra}, {K{\"o}hlinger}, {McFarland}, {Mead}, {Merten}, {Napolitano},
  {Peacock}, {Radovich}, {Schneider}, {Simon}, {Valentijn}, {van den Busch},
  {van Uitert}, \& {Van Waerbeke}}]{kids450}
{Hildebrandt} H. {et~al.}, 2017, \mnras, 465, 1454

\bibitem[{{Hoekstra} \& {Jain}(2008)}]{HoekstraRev}
{Hoekstra} H., {Jain} B., 2008, Annual Review of Nuclear and Particle Science,
  58, 99

\bibitem[{{Jauzac} {et~al}\mbox{.}(2014){Jauzac}, {Cl{\'e}ment}, {Limousin},
  {Richard}, {Jullo}, {Ebeling}, {Atek}, {Kneib}, {Knowles}, {Natarajan},
  {Eckert}, {Egami}, {Massey}, \& {Rexroth}}]{MACSJ0416_HFF}
{Jauzac} M. {et~al.}, 2014, \mnras, 443, 1549

\bibitem[{{Jauzac} {et~al}\mbox{.}(2018){Jauzac}, {Eckert}, {Schaller},
  {Schwinn}, {Massey}, {Bah{\'e}}, {Baugh}, {Barnes}, {Dalla Vecchia},
  {Ebeling}, {Harvey}, {Jullo}, {Kay}, {Kneib}, {Limousin}, {Medezinski},
  {Natarajan}, {Nonino}, {Robertson}, {Tam}, \& {Umetsu}}]{cosmicBeast}
{Jauzac} M. {et~al.}, 2018, \mnras, 481, 2901

\bibitem[{{Jauzac} {et~al}\mbox{.}(2016{\natexlab{a}}){Jauzac}, {Eckert},
  {Schwinn}, {Harvey}, {Baugh}, {Robertson}, {Bose}, {Massey}, {Owers},
  {Ebeling}, {Shan}, {Jullo}, {Kneib}, {Richard}, {Atek}, {Cl{\'e}ment},
  {Egami}, {Israel}, {Knowles}, {Limousin}, {Natarajan}, {Rexroth}, {Taylor},
  \& {Tchernin}}]{substructure_a2744}
{Jauzac} M. {et~al.}, 2016{\natexlab{a}}, \mnras, 463, 3876

\bibitem[{{Jauzac} {et~al}\mbox{.}(2015{\natexlab{a}}){Jauzac}, {Jullo},
  {Eckert}, {Ebeling}, {Richard}, {Limousin}, {Atek}, {Kneib}, {Cl{\'e}ment},
  {Egami}, {Harvey}, {Knowles}, {Massey}, {Natarajan}, {Neichel}, \&
  {Rexroth}}]{harvey_0416}
{Jauzac} M. {et~al.}, 2015{\natexlab{a}}, \mnras, 446, 4132

\bibitem[{{Jauzac} {et~al}\mbox{.}(2012){Jauzac}, {Jullo}, {Kneib}, {Ebeling},
  {Leauthaud}, {Ma}, {Limousin}, {Massey}, \& {Richard}}]{MACSJ0717_HFF}
{Jauzac} M. {et~al.}, 2012, \mnras, 426, 3369

\bibitem[{{Jauzac} {et~al}\mbox{.}(2015{\natexlab{b}}){Jauzac}, {Richard},
  {Jullo}, {Cl{\'e}ment}, {Limousin}, {Kneib}, {Ebeling}, {Natarajan},
  {Rodney}, {Atek}, {Massey}, {Eckert}, {Egami}, \& {Rexroth}}]{A2744_HFF}
{Jauzac} M. {et~al.}, 2015{\natexlab{b}}, \mnras, 452, 1437

\bibitem[{{Jauzac} {et~al}\mbox{.}(2016{\natexlab{b}}){Jauzac}, {Richard},
  {Limousin}, {Knowles}, {Mahler}, {Smith}, {Kneib}, {Jullo}, {Natarajan},
  {Ebeling}, {Atek}, {Cl{\'e}ment}, {Eckert}, {Egami}, {Massey}, \&
  {Rexroth}}]{MACSJ1149_HFF}
{Jauzac} M. {et~al.}, 2016{\natexlab{b}}, \mnras, 457, 2029

\bibitem[{{Jouvel} {et~al}\mbox{.}(2013){Jouvel}, {Host}, {Lahav}, {Seitz},
  {Molino}, {Coe}, {Postman}, {Moustakas}, {Benitez}, {Rosati}, {Balestra},
  {Grillo}, {Bradley}, {Fritz}, {Kelson}, {Koekemoer}, {Lemze}, {Medezinski},
  {Mercurio}, {Moustakas}, {Nonino}, {Scodeggio}, {Zheng}, {Zitrin},
  {Bartelmann}, {Bouwens}, {Broadhurst}, {Donahue}, {Ford}, {Graves},
  {Infante}, {Jimenez-Teja}, {Lazkoz}, {Melchior}, {Meneghetti}, {Merten},
  {Ogaz}, \& {Umetsu}}]{CLASH_photoz}
{Jouvel} S. {et~al.}, 2013, VizieR Online Data Catalog, 356

\bibitem[{{Jullo} {et~al}\mbox{.}(2007){Jullo}, {Kneib}, {Limousin},
  {El{\'{\i}}asd{\'o}ttir}, {Marshall}, \& {Verdugo}}]{lenstool}
{Jullo} E., {Kneib} J.-P., {Limousin} M., {El{\'{\i}}asd{\'o}ttir} {\'A}.,
  {Marshall} P.~J., {Verdugo} T., 2007, New Journal of Physics, 9, 447

\bibitem[{{Kahlhoefer} {et~al}\mbox{.}(2014){Kahlhoefer}, {Schmidt-Hoberg},
  {Frandsen}, \& {Sarkar}}]{SIDMModel}
{Kahlhoefer} F., {Schmidt-Hoberg} K., {Frandsen} M.~T., {Sarkar} S., 2014,
  \mnras, 437, 2865

\bibitem[{{Kilbinger} {et~al}\mbox{.}(2013){Kilbinger}, {Fu}, {Heymans},
  {Simpson}, {Benjamin}, {Erben}, {Harnois-D{\'e}raps}, {Hoekstra},
  {Hildebrandt}, {Kitching}, {Mellier}, {Miller}, {Van Waerbeke}, {Benabed},
  {Bonnett}, {Coupon}, {Hudson}, {Kuijken}, {Rowe}, {Schrabback}, {Semboloni},
  {Vafaei}, \& {Velander}}]{chftpars}
{Kilbinger} M. {et~al.}, 2013, \mnras, 430, 2200

\bibitem[{{Kneib} \& {Natarajan}(2011)}]{massModelReview}
{Kneib} J.-P., {Natarajan} P., 2011, \aapr, 19, 47

\bibitem[{{Koekemoer} {et~al}\mbox{.}(2003){Koekemoer}, {Fruchter}, {Hook}, \&
  {Hack}}]{astrodrizzle}
{Koekemoer} A.~M., {Fruchter} A.~S., {Hook} R.~N., {Hack} W., 2003, in HST
  Calibration Workshop : Hubble after the Installation of the ACS and the
  NICMOS Cooling System, {Arribas} S., {Koekemoer} A., {Whitmore} B., eds., p.
  337

\bibitem[{{Kratochvil} {et~al}\mbox{.}(2010){Kratochvil}, {Haiman}, \&
  {May}}]{peakCosmology}
{Kratochvil} J.~M., {Haiman} Z., {May} M., 2010, \prd, 81, 043519

\bibitem[{{Krist} {et~al}\mbox{.}(2011){Krist}, {Hook}, \& {Stoehr}}]{tinytim}
{Krist} J.~E., {Hook} R.~N., {Stoehr} F., 2011, in Society of Photo-Optical
  Instrumentation Engineers (SPIE) Conference Series, Vol. 8127, Society of
  Photo-Optical Instrumentation Engineers (SPIE) Conference Series

\bibitem[{{Leauthaud} {et~al}\mbox{.}(2007){Leauthaud}, {Massey}, {Kneib},
  {Rhodes}, {Johnston}, {Capak}, {Heymans}, {Ellis}, {Koekemoer}, {Le
  F{\`e}vre}, {Mellier}, {R{\'e}fr{\'e}gier}, {Robin}, {Scoville}, {Tasca},
  {Taylor}, \& {Van Waerbeke}}]{COSMOSintdisp}
{Leauthaud} A. {et~al.}, 2007, \apjs, 172, 219

\bibitem[{{Limousin} {et~al}\mbox{.}(2013){Limousin}, {Morandi}, {Sereno},
  {Meneghetti}, {Ettori}, {Bartelmann}, \& {Verdugo}}]{clusterShapesReview}
{Limousin} M., {Morandi} A., {Sereno} M., {Meneghetti} M., {Ettori} S.,
  {Bartelmann} M., {Verdugo} T., 2013, \ssr, 177, 155

\bibitem[{{Limousin} {et~al}\mbox{.}(2008){Limousin}, {Richard}, {Kneib},
  {Brink}, {Pell{\'o}}, {Jullo}, {Tu}, {Sommer-Larsen}, {Egami},
  {Micha{\l}owski}, {Cabanac}, \& {Stark}}]{a1703}
{Limousin} M. {et~al.}, 2008, \aap, 489, 23

\bibitem[{{Lotz} {et~al}\mbox{.}(2017){Lotz}, {Koekemoer}, {Coe}, {Grogin},
  {Capak}, {Mack}, {Anderson}, {Avila}, {Barker}, {Borncamp}, {Brammer},
  {Durbin}, {Gunning}, {Hilbert}, {Jenkner}, {Khandrika}, {Levay}, {Lucas},
  {MacKenty}, {Ogaz}, {Porterfield}, {Reid}, {Robberto}, {Royle}, {Smith},
  {Storrie-Lombardi}, {Sunnquist}, {Surace}, {Taylor}, {Williams}, {Bullock},
  {Dickinson}, {Finkelstein}, {Natarajan}, {Richard}, {Robertson}, {Tumlinson},
  {Zitrin}, {Flanagan}, {Sembach}, {Soifer}, \& {Mountain}}]{HFF}
{Lotz} J.~M. {et~al.}, 2017, \apj, 837, 97

\bibitem[{{Marian} {et~al}\mbox{.}(2011){Marian}, {Hilbert}, {Smith},
  {Schneider}, \& {Desjacques}}]{peaksnongauss}
{Marian} L., {Hilbert} S., {Smith} R.~E., {Schneider} P., {Desjacques} V.,
  2011, \apjl, 728, L13

\bibitem[{{Massey}(2010)}]{CTI}
{Massey} R., 2010, \mnras, 409, L109

\bibitem[{{Massey} {et~al}\mbox{.}(2010){Massey}, {Kitching}, \&
  {Richard}}]{MKRev}
{Massey} R., {Kitching} T., {Richard} J., 2010, Reports on Progress in Physics,
  73, 086901

\bibitem[{{Massey} {et~al}\mbox{.}(2014){Massey}, {Schrabback}, {Cordes},
  {Marggraf}, {Israel}, {Miller}, {Hall}, {Cropper}, {Prod'homme}, \& {Matias
  Niemi}}]{CTI2}
{Massey} R. {et~al.}, 2014, \mnras, 439, 887

\bibitem[{{McCarthy} {et~al}\mbox{.}(2017){McCarthy}, {Schaye}, {Bird}, \& {Le
  Brun}}]{BAHAMAS}
{McCarthy} I.~G., {Schaye} J., {Bird} S., {Le Brun} A.~M.~C., 2017, \mnras,
  465, 2936

\bibitem[{{Merten} {et~al}\mbox{.}(2009){Merten}, {Cacciato}, {Meneghetti},
  {Mignone}, \& {Bartelmann}}]{MERTEN}
{Merten} J., {Cacciato} M., {Meneghetti} M., {Mignone} C., {Bartelmann} M.,
  2009, \aap, 500, 681

\bibitem[{{Merten} {et~al}\mbox{.}(2015){Merten}, {Meneghetti}, {Postman},
  {Umetsu}, {Zitrin}, {Medezinski}, {Nonino}, {Koekemoer}, {Melchior}, {Gruen},
  {Moustakas}, {Bartelmann}, {Host}, {Donahue}, {Coe}, {Molino}, {Jouvel},
  {Monna}, {Seitz}, {Czakon}, {Lemze}, {Sayers}, {Balestra}, {Rosati},
  {Ben{\'{\i}}tez}, {Biviano}, {Bouwens}, {Bradley}, {Broadhurst}, {Carrasco},
  {Ford}, {Grillo}, {Infante}, {Kelson}, {Lahav}, {Massey}, {Moustakas},
  {Rasia}, {Rhodes}, {Vega}, \& {Zheng}}]{Merten_clash}
{Merten} J. {et~al.}, 2015, \apj, 806, 4

\bibitem[{{Navarro} {et~al}\mbox{.}(1997){Navarro}, {Frenk}, \& {White}}]{NFW}
{Navarro} J.~F., {Frenk} C.~S., {White} S.~D.~M., 1997, \apj, 490, 493

\bibitem[{{Nelson} {et~al}\mbox{.}(2018){Nelson}, {Pillepich}, {Springel},
  {Weinberger}, {Hernquist}, {Pakmor}, {Genel}, {Torrey}, {Vogelsberger},
  {Kauffmann}, {Marinacci}, \& {Naiman}}]{illustrisTNG}
{Nelson} D. {et~al.}, 2018, \mnras, 475, 624

\bibitem[{{Oguri} {et~al}\mbox{.}(2010){Oguri}, {Takada}, {Okabe}, \&
  {Smith}}]{lensingShapes}
{Oguri} M., {Takada} M., {Okabe} N., {Smith} G.~P., 2010, \mnras, 405, 2215

\bibitem[{{Parkinson} {et~al}\mbox{.}(2012){Parkinson}, {Riemer-S{\o}rensen},
  {Blake}, {Poole}, {Davis}, {Brough}, {Colless}, {Contreras}, {Couch},
  {Croom}, {Croton}, {Drinkwater}, {Forster}, {Gilbank}, {Gladders},
  {Glazebrook}, {Jelliffe}, {Jurek}, {Li}, {Madore}, {Martin}, {Pimbblet},
  {Pracy}, {Sharp}, {Wisnioski}, {Woods}, {Wyder}, \& {Yee}}]{wiggles}
{Parkinson} D. {et~al.}, 2012, \prd, 86, 103518

\bibitem[{{Planck Collaboration}(2013)}]{planckpars}
{Planck Collaboration}, 2013, arXiv:1303.5076

\bibitem[{{Postman} {et~al}\mbox{.}(2012){Postman}, {Coe}, {Ben{\'{\i}}tez},
  {Bradley}, {Broadhurst}, {Donahue}, {Ford}, {Graur}, {Graves}, {Jouvel},
  {Koekemoer}, {Lemze}, {Medezinski}, {Molino}, {Moustakas}, {Ogaz}, {Riess},
  {Rodney}, {Rosati}, {Umetsu}, {Zheng}, {Zitrin}, {Bartelmann}, {Bouwens},
  {Czakon}, {Golwala}, {Host}, {Infante}, {Jha}, {Jimenez-Teja}, {Kelson},
  {Lahav}, {Lazkoz}, {Maoz}, {McCully}, {Melchior}, {Meneghetti}, {Merten},
  {Moustakas}, {Nonino}, {Patel}, {Reg{\"o}s}, {Sayers}, {Seitz}, \& {Van der
  Wel}}]{CLASH}
{Postman} M. {et~al.}, 2012, \apjs, 199, 25

\bibitem[{{Ragozzine} {et~al}\mbox{.}(2012){Ragozzine}, {Clowe}, {Markevitch},
  {Gonzalez}, \& {Brada{\v c}}}]{A1758}
{Ragozzine} B., {Clowe} D., {Markevitch} M., {Gonzalez} A.~H., {Brada{\v c}}
  M., 2012, \apj, 744, 94

\bibitem[{{Rasia} {et~al}\mbox{.}(2013){Rasia}, {Meneghetti}, \&
  {Ettori}}]{dynamical_state_xray}
{Rasia} E., {Meneghetti} M., {Ettori} S., 2013, The Astronomical Review, 8, 40

\bibitem[{{Refregier}(2003)}]{RefregierRev}
{Refregier} A., 2003, \araa, 41, 645

\bibitem[{{Rhodes} {et~al}\mbox{.}(2000){Rhodes}, {Refregier}, \&
  {Groth}}]{rrg}
{Rhodes} J., {Refregier} A., {Groth} E.~J., 2000, \apj, 536, 79

\bibitem[{{Rhodes} {et~al}\mbox{.}(2007){Rhodes}, {Massey}, {Albert},
  {Collins}, {Ellis}, {Heymans}, {Gardner}, {Kneib}, {Koekemoer}, {Leauthaud},
  {Mellier}, {Refregier}, {Taylor}, \& {Van Waerbeke}}]{drizzlepars}
{Rhodes} J.~D. {et~al.}, 2007, \apjs, 172, 203

\bibitem[{{Richard} {et~al}\mbox{.}(2010){Richard}, {Smith}, {Kneib}, {Ellis},
  {Sanderson}, {Pei}, {Targett}, {Sand}, {Swinbank}, {Dannerbauer}, {Mazzotta},
  {Limousin}, {Egami}, {Jullo}, {Hamilton-Morris}, \& {Moran}}]{Locuss_Richard}
{Richard} J. {et~al.}, 2010, \mnras, 404, 325

\bibitem[{{Robertson} {et~al}\mbox{.}(2019){Robertson}, Harvey, {Massey},
  {Eke}, {McCarthy}, {Jauzac}, {Li}, \& {Schaye}}]{BAHAMAS_SIDM}
{Robertson} A., Harvey D., {Massey} R., {Eke} V., {McCarthy} I.~G., {Jauzac}
  M., {Li} B., {Schaye} J., 2019, \mnras, 488, 3646

\bibitem[{{S{\'a}nchez} {et~al}\mbox{.}(2012){S{\'a}nchez}, {Sc{\'o}ccola},
  {Ross}, {Percival}, {Manera}, {Montesano}, {Mazzalay}, {Cuesta},
  {Eisenstein}, {Kazin}, {McBride}, {Mehta}, {Montero-Dorta}, {Padmanabhan},
  {Prada}, {Rubi{\~n}o-Mart{\'{\i}}n}, {Tojeiro}, {Xu}, {Maga{\~n}a},
  {Aubourg}, {Bahcall}, {Bailey}, {Bizyaev}, {Bolton}, {Brewington},
  {Brinkmann}, {Brownstein}, {Gott}, {Hamilton}, {Ho}, {Honscheid}, {Labatie},
  {Malanushenko}, {Malanushenko}, {Maraston}, {Muna}, {Nichol}, {Oravetz},
  {Pan}, {Ross}, {Roe}, {Reid}, {Schlegel}, {Shelden}, {Schneider}, {Simmons},
  {Skibba}, {Snedden}, {Thomas}, {Tinker}, {Wake}, {Weaver}, {Weinberg},
  {White}, {Zehavi}, \& {Zhao}}]{sdssboss}
{S{\'a}nchez} A.~G. {et~al.}, 2012, \mnras, 425, 415

\bibitem[{{Schaye} {et~al}\mbox{.}(2015){Schaye}, {Crain}, {Bower}, {Furlong},
  {Schaller}, {Theuns}, {Dalla Vecchia}, {Frenk}, {McCarthy}, {Helly},
  {Jenkins}, {Rosas-Guevara}, {White}, {Baes}, {Booth}, {Camps}, {Navarro},
  {Qu}, {Rahmati}, {Sawala}, {Thomas}, \& {Trayford}}]{eagle}
{Schaye} J. {et~al.}, 2015, \mnras, 446, 521

\bibitem[{{Schwinn} {et~al}\mbox{.}(2018){Schwinn}, {Baugh}, {Jauzac},
  {Bartelmann}, \& {Eckert}}]{substructure_a2744_wavelets}
{Schwinn} J., {Baugh} C.~M., {Jauzac} M., {Bartelmann} M., {Eckert} D., 2018,
  \mnras, 481, 4300

\bibitem[{{Schwinn} {et~al}\mbox{.}(2017){Schwinn}, {Jauzac}, {Baugh},
  {Bartelmann}, {Eckert}, {Harvey}, {Natarajan}, \&
  {Massey}}]{substructure_a2744A}
{Schwinn} J., {Jauzac} M., {Baugh} C.~M., {Bartelmann} M., {Eckert} D.,
  {Harvey} D., {Natarajan} P., {Massey} R., 2017, \mnras, 467, 2913

\bibitem[{{Sereno} {et~al}\mbox{.}(2006){Sereno}, {De Filippis}, {Longo}, \&
  {Bautz}}]{xrayShapes}
{Sereno} M., {De Filippis} E., {Longo} G., {Bautz} M.~W., 2006, \apj, 645, 170

\bibitem[{{Smith} {et~al}\mbox{.}(2005){Smith}, {Kneib}, {Smail}, {Mazzotta},
  {Ebeling}, \& {Czoske}}]{Locuss_Smith}
{Smith} G.~P., {Kneib} J.-P., {Smail} I., {Mazzotta} P., {Ebeling} H., {Czoske}
  O., 2005, \mnras, 359, 417

\bibitem[{{Springel} {et~al}\mbox{.}(2001){Springel}, {White}, {Tormen}, \&
  {Kauffmann}}]{springel01}
{Springel} V., {White} S. D.~M., {Tormen} G., {Kauffmann} G., 2001, \mnras,
  328, 726

\bibitem[{{Teague}(1980)}]{rotatemoments}
{Teague} M.~R., 1980, Journal of the Optical Society of America (1917-1983),
  70, 920

\bibitem[{{Umetsu} {et~al}\mbox{.}(2018){Umetsu}, {Sereno}, {Tam}, {Chiu},
  {Fan}, {Ettori}, {Gruen}, {Okumura}, {Medezinski}, {Donahue}, {Meneghetti},
  {Frye}, {Koekemoer}, {Broadhurst}, {Zitrin}, {Balestra}, {Ben{\'{\i}}tez},
  {Higuchi}, {Melchior}, {Mercurio}, {Merten}, {Molino}, {Nonino}, {Postman},
  {Rosati}, {Sayers}, \& {Seitz}}]{CLASHalignment}
{Umetsu} K. {et~al.}, 2018, \apj, 860, 104

\bibitem[{{Vogelsberger} {et~al}\mbox{.}(2014){Vogelsberger}, {Genel},
  {Springel}, {Torrey}, {Sijacki}, {Xu}, {Snyder}, {Nelson}, \&
  {Hernquist}}]{Illustris}
{Vogelsberger} M. {et~al.}, 2014, \mnras, 444, 1518

\bibitem[{{von der Linden} {et~al}\mbox{.}(2012){von der Linden}, {Allen},
  {Applegate}, {Kelly}, {Allen}, {Ebeling}, {Burchat}, {Burke}, {Donovan},
  {Morris}, {Blandford}, {Erben}, \& {Mantz}}]{WtG}
{von der Linden} A. {et~al.}, 2012, ArXiv e-prints

\bibitem[{{Zitrin} {et~al}\mbox{.}(2015){Zitrin}, {Fabris}, {Merten},
  {Melchior}, {Meneghetti}, {Koekemoer}, {Coe}, {Maturi}, {Bartelmann},
  {Postman}, {Umetsu}, {Seidel}, {Sendra}, {Broadhurst}, {Balestra}, {Biviano},
  {Grillo}, {Mercurio}, {Nonino}, {Rosati}, {Bradley}, {Carrasco}, {Donahue},
  {Ford}, {Frye}, \& {Moustakas}}]{CLASH_zitrin}
{Zitrin} A. {et~al.}, 2015, \apj, 801, 44

\end{thebibliography}

\end{document}